\documentclass[12pt,onecolumn,draftcls]{IEEEtran}

\usepackage{amssymb}
\usepackage{amsmath,epsfig}
\usepackage{subfigure}

\begin{document}

\newtheorem{lemma}{Lemma}

\newcommand{\bo}[1]{\boldsymbol{#1}}
\newcommand{\beq}{\begin{equation}}
\newcommand{\eeq}{\end{equation}}
\newcommand{\etal}{{\it  et al.\ }}
\newcommand{\lb}{\left(}
\newcommand{\rb}{\right)}
\newcommand{\lsb}{\left[}
\newcommand{\rsb}{\right]}
\newcommand{\la}{\left\{ }
\newcommand{\ra}{\right\} }
\newcommand{\lan}{\left\langle }
\newcommand{\ran}{\right\rangle }
\newcommand{\lbg}{\left\lceil}
\newcommand{\rbg}{\right\rceil}
\newcommand{\defi}{\stackrel{\bigtriangleup}{=}}
\newcommand{\Range}{{Range\,}}
\newcommand{\diag}{\,\mbox{diag}\,}
\newcommand{\blockdiag}{\,\mbox{blockdiag}\,}
\newcommand{\diagb}{\,\overline{\mbox{diag}}\,}
\newcommand{\E}{\,\mbox{E}\;}
\newcommand{\MSE}{\mbox{MSE}}
\newcommand{\TS}{N_{TS}}

\newcommand{\dsum}{\displaystyle\sum}
\newcommand{\difrac}{\displaystyle\frac}
\newcommand{\dmin}{\displaystyle\min}
\newcommand{\dmax}{\displaystyle\max}
\newcommand{\dlim}{\displaystyle\lim}
\newcommand{\dinf}{\displaystyle\inf}

\def\adots{\mathinner{\mskip0mu\raise0pt\vbox{\kern7pt\hbox{.}}\mskip3mu
          \raise4pt\hbox{.}\mskip3mu\raise8pt\hbox{.}\mskip0mu}}
\newcommand{\m}{{\!\, -\!\,}}
\newcommand{\p}{{\!\, +\!\,}}
\newcommand{\T}{H}
\newcommand{\tT}{T}
\newcommand{\tcc}{*}
\newcommand{\w}{w}
\newcommand{\z}{{\it z}}        
\newcommand{\eps}{\epsilon}
\newcommand{\oalpha}{\boldsymbol{\alpha}}
\newcommand{\at}{@@}
\newcommand{\pac}{\dagger}      
\newcommand{\Lu}{\underline{L}}
\newcommand{\rank}{\mbox{rank}}
\newcommand{\bfPb}{\overline{\bfP}}

\newcommand{\lla}{l_1}
\newcommand{\llb}{l_2}
\newcommand{\llc}{l_3}
\newcommand{\lld}{l_4}
\newcommand{\lle}{l_5}
\newcommand{\llf}{l_6}
\newcommand{\llg}{l_7}
\newcommand{\llh}{l_8}
\newcommand{\lli}{l_9}
\newcommand{\llj}{l_10}

\newcommand{\bmf}[1]{\mbox{\boldmath ${#1}$}}
\newcommand{\bmfscript}[1]{\mbox{\scriptsize{\boldmath $\displaystyle{#1}$}}}
\newtheorem{Thm}{\bf Theorem}
\newcommand{\mbf}[1]{\mbox{\boldmath ${#1}$}}
\newcommand{\mbfscript}[1]{\mbox{\scriptsize{\boldmath $\displaystyle{#1}$}}}
\newcommand{\bdmA}{\mbox{\boldmath $A$}}
\newcommand{\bdmB}{\mbox{\boldmath $B$}}
\newcommand{\bdmC}{\mbox{\boldmath $C$}}
\newcommand{\bdmD}{\mbox{\boldmath $D$}}
\newcommand{\bdmE}{\mbox{\boldmath $E$}}
\newcommand{\bdmF}{\mbox{\boldmath $F$}}
\newcommand{\bdmG}{\mbox{\boldmath $G$}}
\newcommand{\bdmH}{\mbox{\boldmath $H$}}
\newcommand{\bdmI}{\mbox{\boldmath $I$}}
\newcommand{\bdmJ}{\mbox{\boldmath $J$}}
\newcommand{\bdmK}{\mbox{\boldmath $K$}}
\newcommand{\bdmL}{\mbox{\boldmath $L$}}
\newcommand{\bdmM}{\mbox{\boldmath $M$}}
\newcommand{\bdmN}{\mbox{\boldmath $N$}}
\newcommand{\bdmO}{\mbox{\boldmath $O$}}
\newcommand{\bdmP}{\mbox{\boldmath $P$}}
\newcommand{\bdmQ}{\mbox{\boldmath $Q$}}
\newcommand{\bdmR}{\mbox{\boldmath $R$}}
\newcommand{\bdmS}{\mbox{\boldmath $S$}}
\newcommand{\bdmT}{\mbox{\boldmath $T$}}
\newcommand{\bdmU}{\mbox{\boldmath $U$}}
\newcommand{\bdmV}{\mbox{\boldmath $V$}}
\newcommand{\bdmW}{\mbox{\boldmath $W$}}
\newcommand{\bdmX}{\mbox{\boldmath $X$}}
\newcommand{\bdmY}{\mbox{\boldmath $Y$}}
\newcommand{\bdmZ}{\mbox{\boldmath $Z$}}


\newcommand{\bdma}{\mbox{\boldmath $a$}}
\newcommand{\bdmb}{\mbox{\boldmath $b$}}
\newcommand{\bdmc}{\mbox{\boldmath $c$}}
\newcommand{\bdmd}{\mbox{\boldmath $d$}}
\newcommand{\bdme}{\mbox{\boldmath $e$}}
\newcommand{\bdmf}{\mbox{\boldmath $f$}}
\newcommand{\bdmg}{\mbox{\boldmath $g$}}
\newcommand{\bdmh}{\mbox{\boldmath $h$}}
\newcommand{\bdmi}{\mbox{\boldmath $i$}}
\newcommand{\bdmj}{\mbox{\boldmath $j$}}
\newcommand{\bdmk}{\mbox{\boldmath $k$}}
\newcommand{\bdml}{\mbox{\boldmath $l$}}
\newcommand{\bdmm}{\mbox{\boldmath $m$}}
\newcommand{\bdmn}{\mbox{\boldmath $n$}}
\newcommand{\bdmo}{\mbox{\boldmath $o$}}
\newcommand{\bdmp}{\mbox{\boldmath $p$}}
\newcommand{\bdmq}{\mbox{\boldmath $q$}}
\newcommand{\bdmr}{\mbox{\boldmath $r$}}
\newcommand{\bdms}{\mbox{\boldmath $s$}}
\newcommand{\bdmt}{\mbox{\boldmath $t$}}
\newcommand{\bdmu}{\mbox{\boldmath $u$}}
\newcommand{\bdmv}{\mbox{\boldmath $v$}}
\newcommand{\bdmw}{\mbox{\boldmath $w$}}
\newcommand{\bdmx}{\mbox{\boldmath $x$}}
\newcommand{\bdmy}{\mbox{\boldmath $y$}}
\newcommand{\bdmz}{\mbox{\boldmath $z$}}


\newcommand{\srfA}{\mbox{$\mathsf A$}}
\newcommand{\srfB}{\mbox{$\mathsf B$}}
\newcommand{\srfC}{\mbox{$\mathsf C$}}
\newcommand{\srfD}{\mbox{$\mathsf D$}}
\newcommand{\srfE}{\mbox{$\mathsf E$}}
\newcommand{\srfF}{\mbox{$\mathsf F$}}
\newcommand{\srfG}{\mbox{$\mathsf G$}}
\newcommand{\srfH}{\mbox{$\mathsf H$}}
\newcommand{\srfI}{\mbox{$\mathsf I$}}
\newcommand{\srfJ}{\mbox{$\mathsf J$}}
\newcommand{\srfK}{\mbox{$\mathsf K$}}
\newcommand{\srfL}{\mbox{$\mathsf L$}}
\newcommand{\srfM}{\mbox{$\mathsf M$}}
\newcommand{\srfN}{\mbox{$\mathsf N$}}
\newcommand{\srfO}{\mbox{$\mathsf O$}}
\newcommand{\srfP}{\mbox{$\mathsf P$}}
\newcommand{\srfQ}{\mbox{$\mathsf Q$}}
\newcommand{\srfR}{\mbox{$\mathsf R$}}
\newcommand{\srfS}{\mbox{$\mathsf S$}}
\newcommand{\srfT}{\mbox{$\mathsf T$}}
\newcommand{\srfU}{\mbox{$\mathsf U$}}
\newcommand{\srfV}{\mbox{$\mathsf V$}}
\newcommand{\srfW}{\mbox{$\mathsf W$}}
\newcommand{\srfX}{\mbox{$\mathsf X$}}
\newcommand{\srfY}{\mbox{$\mathsf Y$}}
\newcommand{\srfZ}{\mbox{$\mathsf Z$}}


\newcommand{\srfa}{\mbox{$\mathsf a$}}
\newcommand{\srfb}{\mbox{$\mathsf b$}}
\newcommand{\srfc}{\mbox{$\mathsf c$}}
\newcommand{\srfd}{\mbox{$\mathsf d$}}
\newcommand{\srfe}{\mbox{$\mathsf e$}}
\newcommand{\srff}{\mbox{$\mathsf f$}}
\newcommand{\srfg}{\mbox{$\mathsf g$}}
\newcommand{\srfh}{\mbox{$\mathsf h$}}
\newcommand{\srfi}{\mbox{$\mathsf i$}}
\newcommand{\srfj}{\mbox{$\mathsf j$}}
\newcommand{\srfk}{\mbox{$\mathsf k$}}
\newcommand{\srfl}{\mbox{$\mathsf l$}}
\newcommand{\srfm}{\mbox{$\mathsf m$}}
\newcommand{\srfn}{\mbox{$\mathsf n$}}
\newcommand{\srfo}{\mbox{$\mathsf o$}}
\newcommand{\srfp}{\mbox{$\mathsf p$}}
\newcommand{\srfq}{\mbox{$\mathsf q$}}
\newcommand{\srfr}{\mbox{$\mathsf r$}}
\newcommand{\srfs}{\mbox{$\mathsf s$}}
\newcommand{\srft}{\mbox{$\mathsf t$}}
\newcommand{\srfu}{\mbox{$\mathsf u$}}
\newcommand{\srfv}{\mbox{$\mathsf v$}}
\newcommand{\srfw}{\mbox{$\mathsf w$}}
\newcommand{\srfx}{\mbox{$\mathsf x$}}
\newcommand{\srfy}{\mbox{$\mathsf y$}}
\newcommand{\srfz}{\mbox{$\mathsf z$}}

\newcommand{\oneb}{\mbox{\boldmath $1$}}
\newcommand{\zerb}{\mbox{\boldmath $0$}}

\newcommand{\Us}{\mbox{$\mathcal U$}}
\newcommand{\Vs}{\mbox{$\mathcal V$}}


\newcommand{\bdSA}{\mbox{\bf A}}
\newcommand{\bdSB}{\mbox{\bf B}}
\newcommand{\bdSC}{\mbox{\bf C}}
\newcommand{\bdSD}{\mbox{\bf D}}
\newcommand{\bdSE}{\mbox{\bf E}}
\newcommand{\bdSF}{\mbox{\bf F}}
\newcommand{\bdSG}{\mbox{\bf G}}
\newcommand{\bdSH}{\mbox{\bf H}}
\newcommand{\bdSI}{\mbox{\bf I}}
\newcommand{\bdSJ}{\mbox{\bf J}}
\newcommand{\bdSK}{\mbox{\bf K}}
\newcommand{\bdSL}{\mbox{\bf L}}
\newcommand{\bdSM}{\mbox{\bf M}}
\newcommand{\bdSN}{\mbox{\bf N}}
\newcommand{\bdSO}{\mbox{\bf O}}
\newcommand{\bdSP}{\mbox{\bf P}}
\newcommand{\bdSQ}{\mbox{\bf Q}}
\newcommand{\bdSR}{\mbox{\bf R}}
\newcommand{\bdSS}{\mbox{\bf S}}
\newcommand{\bdST}{\mbox{\bf T}}
\newcommand{\bdSU}{\mbox{\bf U}}
\newcommand{\bdSV}{\mbox{\bf V}}
\newcommand{\bdSW}{\mbox{\bf W}}
\newcommand{\bdSX}{\mbox{\bf X}}
\newcommand{\bdSY}{\mbox{\bf Y}}
\newcommand{\bdSZ}{\mbox{\bf Z}}

\newcommand{\bdSa}{\mbox{\bf a}}
\newcommand{\bdSb}{\mbox{\bf b}}
\newcommand{\bdSc}{\mbox{\bf c}}
\newcommand{\bdSd}{\mbox{\bf d}}
\newcommand{\bdSe}{\mbox{\bf e}}
\newcommand{\bdSf}{\mbox{\bf f}}
\newcommand{\bdSg}{\mbox{\bf g}}
\newcommand{\bdSh}{\mbox{\bf h}}
\newcommand{\bdSi}{\mbox{\bf i}}
\newcommand{\bdSj}{\mbox{\bf j}}
\newcommand{\bdSk}{\mbox{\bf k}}
\newcommand{\bdSl}{\mbox{\bf l}}
\newcommand{\bdSm}{\mbox{\bf m}}
\newcommand{\bdSn}{\mbox{\bf n}}
\newcommand{\bdSo}{\mbox{\bf o}}
\newcommand{\bdSp}{\mbox{\bf p}}
\newcommand{\bdSq}{\mbox{\bf q}}
\newcommand{\bdSr}{\mbox{\bf r}}
\newcommand{\bdSs}{\mbox{\bf s}}
\newcommand{\bdSt}{\mbox{\bf t}}
\newcommand{\bdSu}{\mbox{\bf u}}
\newcommand{\bdSv}{\mbox{\bf v}}
\newcommand{\bdSw}{\mbox{\bf w}}
\newcommand{\bdSx}{\mbox{\bf x}}
\newcommand{\bdSy}{\mbox{\bf y}}
\newcommand{\bdSz}{\mbox{\bf z}}

\newcommand{\cA}{{\mathcal A}}
\newcommand{\cB}{{\mathcal B}}
\newcommand{\cC}{{\mathcal C}}
\newcommand{\cD}{{\mathcal D}}
\newcommand{\cE}{{\mathcal E}}
\newcommand{\cF}{{\mathcal F}}
\newcommand{\cG}{{\mathcal G}}
\newcommand{\cH}{{\mathcal H}}
\newcommand{\cI}{{\mathcal I}}
\newcommand{\cJ}{{\mathcal J}}
\newcommand{\cK}{{\mathcal K}}
\newcommand{\cL}{{\mathcal L}}
\newcommand{\cM}{{\mathcal M}}
\newcommand{\cN}{{\mathcal N}}
\newcommand{\cO}{{\mathcal O}}
\newcommand{\cP}{{\mathcal P}}
\newcommand{\cQ}{{\mathcal Q}}
\newcommand{\cR}{{\mathcal R}}
\newcommand{\cS}{{\mathcal S}}
\newcommand{\cT}{{\mathcal T}}
\newcommand{\cU}{{\mathcal U}}
\newcommand{\cV}{{\mathcal V}}
\newcommand{\cW}{{\mathcal W}}
\newcommand{\cX}{{\mathcal X}}
\newcommand{\cY}{{\mathcal Y}}
\newcommand{\cZ}{{\mathcal Z}}

\newcommand{\overbdmA}{\mbox{\boldmath $\bar{A}$}}
\newcommand{\Tau}{\mbox{$\cT$}}
\newcommand{\overtau}{\mbox{$\bar{\cT}$}}
\newcommand{\tiltau}{\mbox{$\widetilde{\cT}$}}
\newcommand{\overbdmV}{\mbox{\boldmath $\bar{V}$}}

\newcommand{\overdelta}{\mbox{$\bar{\delta}$}}

\newcommand{\sv}{\mbox{$\sigma^2_v$}}
\newcommand{\sa}{\mbox{$\sigma^2_a$}}
\newcommand{\sbb}{\mbox{$\sigma^2_b$}}
\newcommand{\sigs}{\sigma^2}
\newcommand{\sx}{\mbox{$\sigma^2_x$}}

\newcommand{\ah}{\widehat{a}}
\newcommand{\bh}{\widehat{b}}
\newcommand{\Rh}{\widehat{R}}
\newcommand{\Yu}{\underline{Y}}
\newcommand{\Yb}{\overline{Y}}
\newcommand{\Su}{\underline{S}}
\newcommand{\Sba}{\overline{S}}
\newcommand{\bdmcu}{\underline{\bdmc}}
\newcommand{\bdmcb}{\overline{\bdmc}}
\newcommand{\oalphab}{\overline{\oalpha}}

\newcommand{\bdmfh}{\widehat{\bdmf}}

\newcommand{\bdmFt}{\widetilde{\bdmF}}
\newcommand{\bdmGt}{\widetilde{\bdmG}}
\newcommand{\bdmAt}{\widetilde{\bdmA}}

\newcommand{\bfH}{\mbox{\bf H}}
\newcommand{\bfF}{\mbox{\bf F}}
\newcommand{\bfG}{\mbox{\bf G}}
\newcommand{\bfY}{\mbox{\bf Y}}
\newcommand{\bfP}{\mbox{\bf P}}
\newcommand{\bfQ}{\mbox{\bf Q}}
\newcommand{\bfR}{\mbox{\bf R}}
\newcommand{\bfU}{\mbox{\bf U}}
\newcommand{\bfV}{\mbox{\bf V}}
\newcommand{\bfA}{\mbox{\bf A}}
\newcommand{\bfB}{\mbox{\bf B}}
\newcommand{\bfW}{\mbox{\bf W}}
\newcommand{\bfT}{\mbox{\bf T}}
\newcommand{\bfD}{\mbox{\bf D}}
\newcommand{\bfC}{\mbox{\bf C}}
\newcommand{\bfL}{\mbox{\bf L}}
\newcommand{\bfN}{\mbox{\bf N}}
\newcommand{\bfI}{\mbox{\bf I}}
\newcommand{\bfE}{\mbox{\bf E}}
\newcommand{\bfZ}{\mbox{\bf Z}}
\newcommand{\bfX}{\mbox{\bf X}}
\newcommand{\bfM}{\mbox{\bf M}}
\newcommand{\bfO}{\mbox{\bf O}}
\newcommand{\bfJ}{\mbox{\bf J}}

\newcommand{\bfSigma}{\mbox{\boldmath $\Sigma$}}
\newcommand{\bfLambda}{\mbox{\boldmath $\Lambda$}}
\newcommand{\bfPhi}{\mbox{\boldmath $\Phi$}}
\newcommand{\bfalpha}{\mbox{\boldmath $\alpha$}}
\newcommand{\bfDN}{\mbox{\bf DN}}

\newfont{\bb}{msbm10 scaled 1100}
\newcommand{\RR}{\mbox{\bb R}}
\newcommand{\CC}{\mbox{\bb C}}
\newcommand{\FF}{\mbox{\bb F}}

\newcommand{\bfy}{\mbox{\bf y}}
\newcommand{\bfh}{\mbox{\bf h}}
\newcommand{\bff}{\mbox{\bf f}}
\newcommand{\bfv}{\mbox{\bf v}}
\newcommand{\bfx}{\mbox{\bf x}}
\newcommand{\bfa}{\mbox{\bf a}}
\newcommand{\bfb}{\mbox{\bf b}}
\newcommand{\bfr}{\mbox{\bf r}}
\newcommand{\bfe}{\mbox{\bf e}}
\newcommand{\bfw}{\mbox{\bf w}}
\newcommand{\bfc}{\mbox{\bf c}}
\newcommand{\bfu}{\mbox{\bf u}}
\newcommand{\bfg}{\mbox{\bf g}}
\newcommand{\bfs}{\mbox{\bf s}}
\newcommand{\bft}{\mbox{\bf t}}
\newcommand{\bfz}{\mbox{\bf z}}

\newcommand{\bmP}{\mbox{\boldmath $P$}}

\newcommand{\doplus}{\displaystyle\bigoplus}
\newcommand{\Nh}{\widehat{N}}
\newcommand{\bfhh}{\widehat{\bfh}}
\newcommand{\hh}{\widehat{\bfh}}
\newcommand{\bfhhh}{\widehat{\bfhh}}
\newcommand{\htt}{\widetilde{\bfh}}
\newcommand{\oXX}{\overline{\mbox{\bf XX}}}
\newcommand{\oCov}{\overline{Cov}}

\newcommand{\dint}{\displaystyle\int}
\newcommand{\doint}{\displaystyle\oint}
\newcommand{\dprod}{\displaystyle\prod}
\newcommand{\cBb}{\overline{\cB}}
\newcommand{\Cb}{\overline{C}}
\newcommand{\bfS}{\mbox{\bf S}}
\newcommand{\bfHh}{\widehat{\bfH}}
\newcommand{\bfHhh}{\widehat{\bfHh}}
\newcommand{\bfhl}{\overline{\bfh}}
\newcommand{\Hh}{\widehat{\bfH}}
\newcommand{\Ht}{\widetilde{\bfH}}
\newcommand{\Hb}{\overline{\bfH}}
\newcommand{\bfht}{{\widetilde{\bfh}\rule{0mm}{3.4mm}}}
\newcommand{\Vt}{\widetilde{\bfV}}
\newcommand{\bfYb}{\overline{\bfY}}
\newcommand{\bfXb}{\overline{\bfX}}

\newcommand{\bfyh}{\widehat{\bfy}}
\newcommand{\bfyt}{\widetilde{\bfy}}
\newcommand{\Sh}{\widehat{S}}
\newcommand{\St}{\widetilde{S}}
\newcommand{\bfRh}{\widehat{\bfR}}
\newcommand{\bfRt}{\widetilde{\bfR}}
\newcommand{\bfrt}{\overline{\bfr}}

\newcommand{\tr}{\mbox{tr}}
\newcommand{\vect}{\mbox{vec}}
\newcommand{\SNR}{\mbox{SNR}}
\newcommand{\MFB}{\mbox{MFB}}

\newcommand{\xh}{\widehat{x}}
\newcommand{\bfbh}{\widehat{\bfb}}
\newcommand{\bfbt}{\widetilde{\bfb}}
\newcommand{\bt}{\widetilde{b}}
\newcommand{\bfLb}{\overline{\bfL}}
\newcommand{\dbfb}{\Delta{\bfb}}
\newcommand{\dbfc}{\Delta{\bfc}}

\newcommand{\Ns}{N_{s}}
\newcommand{\Nt}{N_{t}}
\newcommand{\Nr}{N_{r}}

\newcommand{\bmh}{\mbox{\boldmath $h$}}

\newcommand{\xb}{\overline{x}}
\newcommand{\yb}{\overline{y}}

\newcommand{\CN}{{\cal CN}}
\newcommand{\SINR}{\mbox{SINR}}
\newcommand{\prob}{\mbox{Prob}}
\newcommand{\bfHb}{\overline{\bfH}}
\newcommand{\bfHt}{\widetilde{\bfH}}
\newcommand{\Lt}{\widetilde{L}}
\newcommand{\Dt}{\widetilde{D}}
\newcommand{\dt}{\widetilde{d}}

\newtheorem{theorem}{Theorem}

\title{Combining Training and Quantized Feedback in Multi-Antenna Reciprocal Channels}
\author{Umer~Salim,~\IEEEmembership{Student Member,~IEEE,}
        David~Gesbert,~\IEEEmembership{Senior Member,~IEEE,}
        and~Dirk~Slock,~\IEEEmembership{Fellow,~IEEE}
\thanks{The authors are with Mobile Communications Department of EURECOM, France
(email: umer.salim@eurecom.fr; david.gesbert@eurecom.fr; dirk.slock@eurecom.fr).
Some part of the material in this paper appears in \cite{u_fb_GC09} to be presented 
at the IEEE Global Communications
Conference (IEEE GLOBECOM), Honolulu, HI, 2009.}}
%
\maketitle
\begin{abstract}
The communication between a multiple-antenna transmitter and multiple receivers (users) with either a single or multiple-antenna each can be significantly enhanced by providing the channel state information at the transmitter (CSIT) of the users, as this allows for scheduling, beamforming and multiuser multiplexing gains. The traditional view on how to enable CSIT has been as follows so far: In time-division duplexed (TDD) systems, uplink (UL) and downlink (DL) channel reciprocity allows the use of a training sequence in the UL direction, which is exploited to obtain an UL channel estimate. This estimate is in turn recycled in the next downlink slot. In frequency-division duplexed (FDD) systems, which lack the UL and DL reciprocity, the CSIT is provided via the use of a dedicated feedback link of limited capacity between the receivers and the transmitter. In this paper, we focus on TDD systems and put this classical approach in question. In particular, we show that the traditional TDD setup above fails to fully exploit the channel reciprocity in its true sense. In fact, we show that the system can benefit from a combined CSIT acquisition strategy mixing the use of limited feedback and that of a training sequence. This combining gives rise to a very interesting joint estimation and detection problem for which we propose two iterative algorithms. An outage rate based framework is also developed which gives the optimal resource split between training and feedback. We demonstrate the potential of this hybrid combining in terms of the improved CSIT quality under a global training and feedback resource constraint.
\end{abstract}                  
\section{Introduction}
\label{sec:intro}
Multiple-antenna transmitters and receivers are instrumental to optimizing the performance of bandwidth and power limited wireless communication systems. In the downlink (DL), in particular, the communication between a multiple-antenna enabled base station (BS) and one or more users with either a single or multiple antenna each can be significantly enhanced through the use of scheduling, beamforming and power allocation algorithms, be it in single user or multi-user mode (spatial division multiplexing). To allow for beamforming and/or multi-user multiplexing capability, the BS transmitter must however be informed with the channel state information (CSI) of each of the served users \cite{caire_bc}, \cite{gesbert_mu_mimo}, except when the number of users reaches an asymptotic (large) regime in which case random opportunistic beamforming scheme can be exploited \cite{tse_dumb_antennas}, \cite{hassibi_RBF}. This has motivated the proposal of many techniques for providing the channel state information at the transmitter (CSIT) in an efficient manner. Proposals for how to provide CSIT roughly fall in two categories depending upon the chosen duplexing scheme for the considered wireless network. In the case of time-division duplex (TDD) systems, it was always assumed that CSIT should exploit the reciprocity of the uplink (UL) and DL channels, so as to avoid the use of any resource consuming feedback channel \cite{marz_MU_Training}, \cite{marz_csi_transfer}. The way reciprocity is exploited in the current TDD systems, is through the use of a training sequence sent by the user on the UL, based on which the BS first builds an estimate of the UL channel which in turn serves as an estimate for the DL channel in the next DL slot \cite{marz_MU_Training}. In frequency-division duplex (FDD) systems, UL and DL portions of the bandwidth are normally quite apart and hence the channel realizations can be safely assumed to be independent of each other. This lack of channel reciprocity motivates instead the use of a dedicated feedback link in which the user conveys the information, about the estimated DL channel, back to the BS. Recently, several interesting strategies have been proposed for how to best use a limited feedback channel and still provide the BS with exploitable CSIT (see \cite{love_fb_value}, \cite{jindal_fb06}, \cite{yoo_ZF_Scheduling}, \cite{love_Overview_FB} and the references therein for further details).

Although in the past, the balance has weighed in the favor of FDD systems when choosing a duplexing scheme (in part because of heavy legacy issues in voice oriented 2G networks and also because of interference management between UL and DL), current discussions in the standardization groups indicate an increasing level of interest for TDD for upcoming wireless data-access networks (e.g.WiMax, etc.), caused partly by its advantages in maintaining system flexibility with respect to UL and DL traffic loads, and mostly because TDD systems are seen as more efficient in providing the CSIT required by several MIMO DL schemes, thanks to the channel reciprocity.

In this paper, we focus on the problem of CSIT acquisition in a TDD system. We take a step back and shed some critical light on the traditional approach above consisting in exploiting the channel reciprocity via the use of training sequences exclusively. In fact we show that this approach fails to fully exploit the channel reciprocity. The key shortcoming is as follows: when sending a training sequence in the UL of a traditional TDD system, the user allows the BS to estimate the channel by a classical channel estimator (it can be a least-square (LS) estimator or minimum mean square error (MMSE) based, just to name a few). However, note that the user itself has the knowledge of the channel coefficients (obtained during the current DL frame or from the DL synchronization sequence or other control signals or even from the previous DL frames if the channel is correlated in time) but, regretfully, does not exploit that knowledge in order to facilitate the CSIT acquisition by the BS. Instead, it uses this knowledge only locally. 

Interestingly, by contrast, in FDD systems, the user exploits its DL channel knowledge by quantizing the channel and sending the result over a dedicated feedback link (actually UL bandwidth is used for this feedback along with UL data transmission). In the FDD case, UL training is used by the BS solely for UL data detection as this UL training cannot give any direct information to the BS about the DL channel coefficients. 

In this paper, we point out that in TDD systems there is a unique opportunity to combine the advantages of both forms of CSIT acquisition. In doing so, we obtain a new CSIT acquisition scheme mixing the classical channel estimation using training with the quantized limited channel feedback of the same channel. This gives us a framework for fully utilizing the channel reciprocity in a TDD setup and it improves the classical trade-off between the CSIT quality and the amount of training/feedback resource used. We characterize the optimal CSIT acquisition structure under this novel framework. This hybrid CSIT acquisition setup gives rise to a very interesting joint estimation and detection problem for which we propose two iterative algorithms. We further propose a sub-optimal outage rate based approach which helps us to optimize the fixed resource partitioning between training and quantized feedback phases. We adapt this optimization framework to use it with practical constellations like QSPK and 16-QAM. The results obtained confirm our intuition and clearly demonstrate the benefit of this hybrid (mix of training and quantized feedback) approach for upcoming TDD systems. 

In previous work, Caire et al. studied the achievable rates for multi-user MIMO DL removing all the assumptions of channel state information at the receiver (CSIR) and CSIT for FDD systems in \cite{caire_mu_mimo}. They gave transmission schemes incorporating all the necessary training and feedback stages and compared achievable rates either with analog feedback or with quantized feedback. The reference \cite{Tejera_ICC07} studies the decay rate of the feedback distortion versus SNR with analog and digital quantized feedback for FDD systems. A very recent paper \cite{caire_fb_mimo_mac_ISIT09} studies combining the analog and digital feedback for FDD systems. All of these works fundamentally differ from our work as there is no channel reciprocity in FDD systems and hence there is no point in combining the UL training and the quantized feedback of the DL channel.

Some other contributions \cite{marz_MU_Training}, \cite{Jose_ICC08}, \cite{u_allerton08}, \cite{Jose_arxiv1208} and \cite{u_ISWCS09} analyze the sum rate of TDD systems starting without any assumption of CSI but restrict the CSIT acquisition through training only. \cite{marz_csi_transfer} does a comparison of TDD systems versus FDD systems in terms of CSIT acquisition accuracy. \cite{Sabharwal_ITWC08} studies the diversity-multiplexing trade-off \cite{Tse_DMT} of two-way SIMO channels when TDD is the mode of operation. All of these references treat no-CSI TDD systems but all acquire CSIT through training only. According to authors' knowledge, there is no single contribution which exploits the combining of training and the quantized feedback in TDD systems, which we believe to be one of the major novelties of this work.

The paper is structured as follows: The system model is given in section \ref{sec:model}, followed by the classical CSIT acquisition for FDD and TDD systems in section \ref{sec:CSIT_Acq}. The optimal CSIT acquisition strategy combining training and feedback is outlined in section \ref{sec:combining}. Two iterative and one non-iterative algorithms for the joint estimation and detection have been proposed in section \ref{sec:algos}. The simplified outage-rate based framework to optimize the resource split appears in section \ref{sec:outage_based_opt} followed by its adaption for practical constellations in section \ref{sec:constellation_opt}. The simulation results have been provided in section \ref{sec:results}, followed by the conclusions and the possible future extensions combined in section \ref{sec:conc}.\\
\textbf{\textit{Notation:}} $\mathbb{E}$ denotes statistical expectation. Lowercase letters represent scalars, boldface lowercase letters represent vectors, and boldface uppercase letters denote matrices. $\mathbf{A^\dagger}$ and $\mathbf{A^{-1}}$ denote the Hermitian and the inverse of matrix $\mathbf{A}$, respectively. For a vector $\mathbf{a}$, $\mathbf{||a||}$ and $\mathbf{\bar{a}}$ represent, respectively, its norm and unit-norm direction vector so that $\mathbf{a = ||a|| \bar{a}}$. A Gaussian distributed vector $\mathbf{a}$ with mean $\mathbf{m_a}$ and covariance matrix $\mathbf{K_a}$ is represented as $\mathbf{a} \sim \mathcal{CN}\left(\mathbf{m_a},\mathbf{K_a}\right)$. $\mathbf{I_M}$ represents the identity matrix of $M$ dimensions.

\section {System Model and CSIR Acquisition}
\label{sec:model}
We consider the two way communication in a cell between a single BS, equipped with $M$ antennas, and a single antenna mobile user. The DL channel $\mathbf{h} \in \mathbb{C}^M$ is assumed to be flat-fading with independent complex Gaussian zero-mean unit-variance entries, where $\mathbb{C}^M$ represents the $M$-dimensional complex space. We assume block fading channel so each channel realization stays constant for $T$ channel uses \cite{marz} which can be accordingly partitioned between UL and DL data transmissions. 

The goal of this work is to provide a reliable estimate of the DL channel to the BS, which in turn can be used for beamforming/precoding purposes. However we focus on the acquisition issue of the channel knowledge and not about its use in MIMO transmission schemes.

In the downlink, the received signal at the user for $L$ symbol intervals is given by
\beq
\mathbf{y_{dl} = X_{dl} h + n_{dl}},
\eeq
where $\mathbf{X_{dl}} \in \mathbb{C}^{L \times M}$ is the signal transmitted by the BS for $L$ channel uses (satisfying BS power constraint), $\mathbf{n_{dl}}\in \mathbb{C}^{L}$ is the complex Gaussian noise with independent zero-mean unit-variance entries and $\mathbf{y_{dl}}\in \mathbb{C}^{L}$ is the observation sequence during this $L$-length interval. 

If we want to use the above DL system equation for channel estimation, for identifiability of $M$-dimensional channel at the user's side, the length of the transmitted data (the training sequence in this case) should be larger than $M$, the number of BS transmit antennas. Based upon the knowledge of the transmitted data $\mathbf{X_{dl}}$ (the training sequence) and the observed sequence $\mathbf{y_{dl}}$, the user can estimate the DL channel $\mathbf{h}$ using various techniques. The LS estimate, denoted as $\mathbf{\check{h}_{LS}}$, would be \cite{hassibi}
\beq
\mathbf{\check{h}_{LS} = \left(X_{dl}^{\dagger}  X_{dl}\right)^{-1} X_{dl}^{\dagger}  y_{dl}}.
\eeq
The user can make a better channel estimate using MMSE criteria, and the estimate is given by
\beq
\mathbf{\check{h}_{MMSE} = \left(X_{dl}^{\dagger} X_{dl} + I_M\right)^{-1} X_{dl}^{\dagger} y_{dl} }.
\eeq
%
\section {Classical CSIT Acquisition in FDD and TDD}
\label{sec:CSIT_Acq}
We now briefly review the classical approaches for acquiring CSIT at the BS in FDD and TDD systems. We shall build upon the equations below in order to present our ideas later. 
\subsection{FDD Systems}
\begin{figure}[htbp]
	\begin{center}
		\includegraphics[scale=0.6]{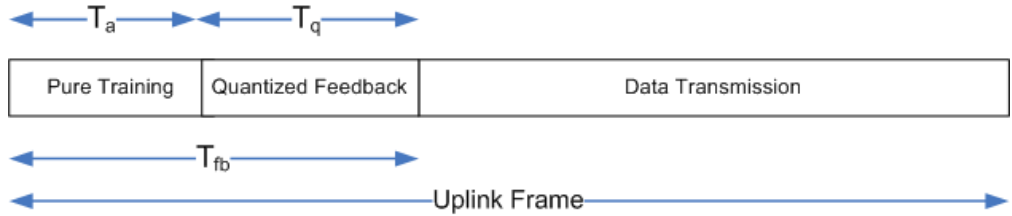}
	\end{center}
	\caption{Uplink frame structure: Total feedback length is divided between UL training and quantized feedback phases.}
	\label{fig:frame}
\end{figure}

A typical UL frame for FDD systems is shown in Fig. (\ref{fig:frame}) where the initial $T_{fb}$ channel uses are reserved for feedback.

For the BS to be able to decode the feedback properly (sent as UL payload), it should first know/estimate the UL channel (denoted as $\mathbf{h_u} \in \mathbb{C}^M$). If the user sends a normalized training sequence $\mathbf{x_a} \in \mathbb{C}^{1 \times T_a}$ of length $T_a$ in the UL direction, the signal received at the BS for $T_a$ channel uses is given by
\beq
\mathbf{Y_{a}} = \sqrt{P} \; \mathbf{h_u x_a + N_a},
\label{eq:FDD_analog}
\eeq
where $\mathbf{N_a} \in \mathbb{C}^{M \times T_a}$ represents the spatio-temporally white Gaussian noise with zero-mean unit-variance entries and $\mathbf{Y_{a}} \in \mathbb{C}^{M \times T_a}$ is the received signal at $M$ antennas of the BS during this $T_a$-length training interval. $P$ represents the user's peak power constraint which is equal to the UL signal-to-noise-ratio (SNR) at every BS antenna due to the normalized noise variances. After observing $\mathbf{Y_{a}}$, the BS can make an estimate $\mathbf{\hat{h}_u}$ of the UL channel $\mathbf{h_u}$, knowing the training sequence $\mathbf{x_a}$. Estimation techniques like LS or MMSE as described in the previous section can be applied.

In FDD systems, the mobile station obtains the DL channel estimate $\mathbf{\check{h}}$ from the DL frame as described in the previous section. If $Q$ denotes the quantization function, then for the DL channel estimate $\mathbf{\check{h}}$, its quantized version (the index of the closest codeword in the codebook) is given by $Q(\mathbf{\check{h}})$. Afterward user maps this index (sequence of bits) into a sequence of constellation symbols, using the mapping function denoted by $S$. Let the finite cardinality set of all mapped codewords be denoted by $\mathcal{CB}$. Hence the feedback of the DL channel would be
\beq
\mathbf{x_{q}} = S(Q(\mathbf{\check{h}})),
\eeq
where $\mathbf{x_{q}} \in \mathbb{C}^{1 \times T_q}$ is the $T_q$ dimensional row vector of the normalized constellation symbols. The signal received at the BS upon transmission of $\mathbf{x_{q}}$ is
\beq
\mathbf{Y_{q}} = \sqrt{P} \; \mathbf{h_u x_{q} + N_{q}},
\label{eq:FDD_digital}
\eeq
where $\mathbf{Y_{q}}$ and $\mathbf{N_{q}}$ are $M \times T_q$ matrices of the received signal and the noise respectively at $M$ antennas of the BS during this explicit $T_q$ length feedback interval. So based upon the estimate $\mathbf{\hat{h}_u}$ of the UL channel $\mathbf{h_u}$ and the received feedback $\mathbf{Y_{q}}$, BS tries to recover the DL channel feedback (quantized version, $\mathbf{x_{q}}$) using the optimum (although relatively complex) maximum likelihood (ML) sequence estimation technique.
\beq
\mathbf{\hat{h}} = \underset{\mathbf{\check{h}}}{\operatorname{arg\,min}} \; || \mathbf{Y_{q}} -  \sqrt{P} \; \mathbf{\hat{h}_u} S(Q(\mathbf{\check{h}}))||^2
\eeq
The search space will be restricted to the codebook, hence the BS, at best, can estimate the quantized version of the channel.
\subsection{TDD Systems}
If the communication system is operating under TDD mode, DL and UL channels are reciprocal, hence $\mathbf{h_u = h}$. So if a user transmits pilot sequence on the UL (like eq. (\ref{eq:FDD_analog})), the simple (UL) channel estimation at the BS furnishes CSIT due to UL and DL channel reciprocity. In the past, this has been the classical way of getting CSIT in TDD systems \cite{marz_MU_Training}, \cite{marz_csi_transfer}. 
\section{Optimal Training and Feedback Combining in TDD Systems}
\label{sec:combining}
The classical training based CSIT acquisition for TDD systems ignores the fact that user knows the DL channel and the CSIT acquisition based only on the quantized feedback for FDD systems cannot use the channel reciprocity whereas in TDD systems both can be exploited at the same time.  

We propose a novel hybrid two stage CSIT acquisition strategy which exploits the channel reciprocity and user's channel knowledge at the same time. We assume perfect channel knowledge at the user's side for ease of exposition\footnote{In general, the CSIR quality at the users' side is much better. Firstly the DL pilots are global (they are not transmitted per user contrary to the UL pilots) and secondly, the BS can surely pump larger power as compared to small hand-held mobile devices.} and later, in section \ref{sec:results_imperfect_CSIR}, we show results removing this perfect CSIR assumption. Working under a constraint of fixed resource available for CSIT acquisition ($T_{fb}$ channel uses and user's power constraint of $P$), our strategy consists of dividing this interval in two phases as shown in Fig. (\ref{fig:frame}), contrary to the classical pilot sequence transmission. The first stage of this hybrid approach, termed as ``pure training", is the transmission of training sequence from the user to the BS for $T_a$ channel uses and the received signal will be
\beq
\mathbf{Y_{a}} = \sqrt{P} \;  \mathbf{h x_a + N_a}.
\label{eq:TDD_analog}
\eeq
(See eq. (\ref{eq:FDD_analog}) for the dimensions of all parameters.)\\
The optimal training based estimate, denoted as $\mathbf{\hat{h}_a}$, based upon the observed signal $\mathbf{Y_a}$ and knowing $\mathbf{x_a}$ will be
\beq
\mathbf{\hat{h}_a} = \underset{\mathbf{h}}{\operatorname{arg\,min}} \; || \mathbf{Y_a} - \sqrt{P} \; \mathbf{h x_a}||^2
\label{eq:analog_CSIT}
\eeq
The second stage, termed as ``quantized feedback", consists of the transmission of quantized channel, already known at the user, for $T_q$ channel uses and the received signal will be 
\beq
\mathbf{Y_{q}} = \sqrt{P} \; \mathbf{h x_{q} + N_{q}},
\label{eq:TDD_digital}
\eeq
(See eq. (\ref{eq:FDD_digital}) for the dimensions of all parameters.)\\
where $\mathbf{x_{q}} = S(Q(\mathbf{h})) \in \mathcal{CB}$. This equation reveals the intriguing aspect that the BS needs to acquire $\mathbf{h}$ which appears both as the channel and the transmitted feedback $\mathbf{x_{q}}$. The BS can try to decode only the quantized channel information based upon the knowledge of $\mathbf{\hat{h}_a}$ (obtained as in eq. (\ref{eq:analog_CSIT}) making use of pure training $\mathbf{x_{a}}$)
\beq
\mathbf{\hat{h}_q} = \underset{\mathbf{x_{q}} \in \mathcal{CB}}{\operatorname{arg\,min}} \; || \mathbf{Y_{q}} - \sqrt{P} \; \mathbf{\hat{h}_a x_{q}}||^2.
\label{eq:digital_CSIT}
\eeq

The optimal CSIT will be obtained by the joint estimation and detection (of $\mathbf{h}$ and $\mathbf{x_{q}}$ respectively) based upon the observation of $\mathbf{Y_a}$ and $\mathbf{Y_{q}}$, knowing $\mathbf{x_a}$ and assuming an optimal split between the training and the quantized feedback phases (constrained as $T_a + T_q = T_{fb}$).
\beq
\mathbf{\hat{h}} = \underset{\mathbf{h}}{\operatorname{arg\,min}} \; || \; \mathbf{[Y_a \; Y_{q}}] - \sqrt{P} \; \mathbf{h [x_a} \; S(Q(\mathbf{h}))] \; ||^2
\label{eq:optimal_CSIT}
\eeq
The optimal solution requires a double minimization and does not seem to bear a closed form expression for $\mathbf{\hat{h}}$. 
\section{Algorithms for Joint Channel Estimation and Feedback Detection}
\label{sec:algos}
We give three algorithms in this section which separately solve the estimation and the detection problem of the joint minimization of eq. (\ref{eq:optimal_CSIT}). The first two algorithms are iterative which separately solve the estimation and detection problems and iterate till convergence. These algorithms have been closely inspired by \cite{Talwar_ITSP96} which proposes similar algorithms for joint blind estimation and detection for signal separation. We have made modifications for our requirements where data aided channel estimation after the initialization step and the presence of channel as ``data" (feedback) give it its unique texture. The third algorithm is just the single-shot solution of the joint estimation and detection. Owing to its simplicity, it allows us to further optimize the resource split between training and quantized feedback in the next section.

\subsection{Iterative Estimation and Detection}
We describe below our algorithm.\\
Step 1)\; Initial channel estimation based only upon pilots
\beq
\mathbf{\hat{h}_a}^0 = \underset{\mathbf{h}}{\operatorname{arg\,min}} \; || \mathbf{Y_a} - \sqrt{P} \; \mathbf{h x_a}||^2,
\eeq
which is a simple least squares problem with the solution
\beq
\mathbf{\hat{h}_a}^0 = \mathbf{Y_a x_a^{\dagger} (x_a x_a^{\dagger})^{-1} } \frac{1}{\sqrt{P}}.
\eeq
\beq
i=1
\eeq
Superscript denotes the iteration number.\\
Step 2)\; At iteration $i$, do enumeration over all the codes in the codebook assuming that the channel $\mathbf{\hat{h}_a}^{i-1}$ is perfectly known.
\beq
\mathbf{\hat{x}_q}^i = \underset{\mathbf{x_{q}} \in \mathcal{CB}}{\operatorname{arg\,min}} \; || \mathbf{Y_{q}} - \sqrt{P} \; \mathbf{\hat{h}_a}^{i-1} \mathbf{x_{q}}||^2
\eeq
Step 3)\; Regenerate extended pilot sequence $\mathbf{x_{ext}}$ (pilots and detected feedback)
\beq
\mathbf{x_{ext}}^i =  [\mathbf{x_a} \; \mathbf{\hat{x}_q}^i]. \hspace{1cm} \mathbf{Y_{ext}} = \mathbf{[Y_a \; Y_q]}.
\eeq
Step 4)\; Channel estimation based upon extended pilots (i.e. knowing $\mathbf{x_{ext}}^i$)
\beq
\mathbf{\hat{h}_a}^i = \underset{\mathbf{h}}{\operatorname{arg\,min}} \; || \mathbf{Y_{ext}} - \sqrt{P} \; \mathbf{h x_{ext}}^i||^2
\eeq
\beq
\mathbf{\hat{h}_a}^i = \mathbf{Y_{ext} x_{ext}}^{\dagger i} (\mathbf{x_{ext}}^i \mathbf{x_{ext}}^{\dagger i})^{-1}  \frac{1}{\sqrt{P}} 
\eeq
Step 5)\; If $\mathbf{\hat{x}_q}^i \neq \mathbf{\hat{x}_q}^{i-1}$ or $\mathbf{\hat{h}_a}^i \neq \mathbf{\hat{h}_a}^{i-1}$, $i = i+1$ and go to Step $2$.\\
The final channel estimate $\mathbf{\hat{h}}$ is the channel vector corresponding to $\mathbf{\hat{x}_q}^i$ in the codebook.\\
\begin{theorem}[Convergence for Iterative Estimation and Detection Algorithm]
Let $\mathbf{\hat{h}_a}^i$ be the estimated channel and $\mathbf{\hat{x}_q}^i$ be the detected feedback, both at $i$-th iteration of the iterative estimation and detection algorithm. Let the residual function $f\left(\mathbf{\hat{h}_a},\mathbf{x_{ext}};\mathbf{Y_{ext}}\right) \stackrel{\Delta}{=} || \mathbf{Y_{ext}} - \sqrt{P} \; \mathbf{\hat{h}_a x_{ext}}||^2$ be selected as the descent function for this algorithm. Then there exists some $j$ such that for any $i \geq j$, $\mathbf{\hat{x}_q}^i = \mathbf{\hat{x}_q}^j$ and $\mathbf{\hat{h}_a}^i = \mathbf{\hat{h}_a}^j$.
\label{th:convergence_IEDA}
\end{theorem}
\begin{proof}
The residual descent function $f\left(\mathbf{\hat{h}_a},\mathbf{x_{ext}};\mathbf{Y_{ext}}\right) = || \mathbf{Y_{ext}} - \sqrt{P} \; \mathbf{\hat{h}_a x_{ext}}||^2$ is clearly non-negative and continuous. Considering the residual function at $i$-th iteration:
\begin{eqnarray}
f\left(\mathbf{\hat{h}_a}^i,\mathbf{x_{ext}}^i;\mathbf{Y_{ext}}\right) & \stackrel{a}{=} & || \mathbf{Y_{ext}} - \sqrt{P} \; \mathbf{\hat{h}_a}^i \mathbf{x_{ext}}^i||^2 \nonumber \\ 
 & \stackrel{b}{=}  & \underset{\mathbf{h}}{\operatorname{min}} \; || \mathbf{Y_{ext}} - \sqrt{P} \; \mathbf{h x_{ext}}^i||^2 \nonumber \\
 & \stackrel{c}{\leq} & || \mathbf{Y_{ext}} - \sqrt{P} \; \mathbf{\hat{h}_a}^{i-1} \mathbf{x_{ext}}^i||^2 \nonumber \\
 & \stackrel{d}{=} & || \mathbf{Y_{a}} - \sqrt{P} \; \mathbf{\hat{h}_a}^{i-1} \mathbf{x_{a}}||^2 + || \mathbf{Y_{q}} - \sqrt{P} \; \mathbf{\hat{h}_a}^{i-1} \mathbf{\hat{x}_q}^i||^2 \nonumber \\
 & \stackrel{e}{=} & || \mathbf{Y_{a}} - \sqrt{P} \; \mathbf{\hat{h}_a}^{i-1} \mathbf{x_{a}}||^2 +  \underset{\mathbf{x_{q}} \in \mathcal{CB}}{\operatorname{min}} \; || \mathbf{Y_{q}} - \sqrt{P} \; \mathbf{\hat{h}_a}^{i-1} \mathbf{x_{q}}||^2 \nonumber \\
 & \stackrel{f}{\leq} & || \mathbf{Y_{a}} - \sqrt{P} \; \mathbf{\hat{h}_a}^{i-1} \mathbf{x_{a}}||^2 +  || \mathbf{Y_{q}} - \sqrt{P} \; \mathbf{\hat{h}_a}^{i-1} \mathbf{\hat{x}_{q}}^{i-1}||^2 \nonumber \\
 & \stackrel{g}{=} & || \mathbf{Y_{ext}} - \sqrt{P} \; \mathbf{\hat{h}_a}^{i-1} \mathbf{x_{ext}}^{i-1}||^2 \nonumber \\
 & \stackrel{h}{=} & f\left(\mathbf{\hat{h}_a}^{i-1},\mathbf{x_{ext}}^{i-1};\mathbf{Y_{ext}}\right)
\end{eqnarray}
Equalities $d$ and $g$ make use of the property of the Frobenius norm \cite{horn_MA}. The set of equations above shows that each single iteration of the algorithm over estimation and detection causes to monotonically reduce the residual function unless iterates converge. This monotonic reduction of the descent function, its non-negativity and the fact that $\mathbf{x_{q}}$ belongs to a finite set (codes of the codebook) and hence corresponding iterates of the estimation subproblem are also finite prove the convergence of this algorithm to the locally optimal solution in a finite number of steps. The globally optimal solution is achieved by having a good initial point which depends upon the training part as confirmed by our simulations.
\end{proof}
\subsection{Simplified Iterative Estimation and Detection}
This algorithm is very similar to the previous algorithm in essence but the difference arises at the detection step. The second step of the previous algorithm, the ML detection of the quantized code from the codebook, is computationally quite onerous, especially for codebooks with large cardinality. So we replace this enumeration step with least squares detection followed by mapping on the codebook. So the Step $2$ of the previous algorithm gets replaced by two sub-steps.\\
Step 2-A)\; At iteration $i$, do LS detection of the quantized feedback assuming $\mathbf{\hat{h}_a}^{i-1}$ as the perfectly known channel
\beq
\mathbf{\hat{x}_{\mathrm{LS}}}^i =  (\mathbf{\hat{h}_a}^{\dagger i-1} \mathbf{\hat{h}_a}^{i-1})^{-1} \mathbf{\hat{h}_a}^{\dagger i-1} \mathbf{Y_{q} }\frac{1}{\sqrt{P}}.
\eeq
Step 2-B)\; Do hard detection on the constellation symbols which will map the LS channel estimate to the nearest code in the codebook.
\beq
\mathbf{\hat{x}_q}^i =  \mathrm{Hard Detection}(\mathbf{\hat{x}_{\mathrm{LS}}}^i)
\eeq

This helps to significantly reduce the computational complexity. Later results show that this does not involve any discernible performance degradation.

\subsection{Single-Shot Estimation and Detection}
This is the simplest and the fastest algorithm for the joint estimation and detection problem where the channel estimation and the feedback detection are performed (separately) only once.\\
Step 1)\; Channel estimation based only upon the pilots
\beq
\mathbf{\hat{h}_a} = \underset{\mathbf{h}}{\operatorname{arg\,min}} \; || \mathbf{Y_a} - \sqrt{P} \; \mathbf{h x_a}||^2.
\eeq
We can employ either the LS or the MMSE estimation technique. \\
Step 2)\; Detection of the feedback $\mathbf{x_{q}}$ assuming channel $\mathbf{\hat{h}_a}$ is perfectly known. This detection problem can be solved either by enumerating all the codewords like the first algorithm or by simple LS like the second algorithm or even by applying MMSE filter.

\section{Outage Based Training and Feedback Partitioning}
\label{sec:outage_based_opt}
\subsection{Definitions and Initial Setup}
The solution for the optimal CSIT estimate, $\mathbf{\hat{h}}$ in eq. (\ref{eq:optimal_CSIT}), requires joint estimation and detection. Furthermore, the fixed resource ($T_{fb}$ channel uses) needs to be optimally split between training and feedback. Even if, as a simplification, we focus separately on training based estimate $\mathbf{\hat{h}_a}$ (given in eq. (\ref{eq:analog_CSIT})) and digital feedback based estimate $\mathbf{\hat{h}_q}$ (given in eq. (\ref{eq:digital_CSIT})), two questions arise: i) how the fixed CSIT acquisition interval $T_{fb}$ should be split between training and feedback?, and ii) how the two estimates should be combined to get the final estimate?

We use the minimization of the mean-square error (MSE) of the final CSIT (defined below) as the criterion for the optimal resource split, thus answering the first question for which we give the proper framework in the next subsection. It has been shown in \cite{jindal_fb06} that the principal factor in the DL sum rate loss due to imperfect CSI is the MSE of CSIT. Hence the minimization of the MSE of CSIT is equivalent to the maximization of the system wide sum rate, the most commonly adopted performance metric. Furthermore, we propose to use the quantized feedback based estimate $\mathbf{\hat{h}_q}$ as the final CSIT estimate $\mathbf{\hat{h}}$ due to better channel diversity exploitation properties of digital transmission as an answer to the second question. It may give the impression that the training based estimate $\mathbf{\hat{h}_a}$ goes wasted but in reality quantized feedback $\mathbf{x_{q}}$, which provides $\mathbf{\hat{h}_q}$, is decoded based upon this training based estimate $\mathbf{\hat{h}_a}$.


This optimization framework consists of first providing a training based estimate $\mathbf{\hat{h}_a}$ to the BS in the training interval of $T_a$ channel uses. In the second interval of $T_q$ channel uses, the user sends the quantized version of its unit-norm channel direction information (CDI) vector which we assume to be perfectly known at the user. As the channel stays constant for each acquisition interval, this feedback transmission is equivalent to the transmission over slow fading channels for which deep channel fades (causing outage) are the typical error events \cite{tse_book}. We define the ``outage" as an event when the channel realization and the quality of the training based estimate $\mathbf{\hat{h}_a}$ (a function of $T_a$) don't allow the BS to successfully decode the feedback information. Let $\epsilon(T_a,b)$ be the outage probability when transmitting $b$ bits per channel use on the UL feedback channel. Thus $b$ is the $\epsilon(T_a,b)$-outage rate of the UL channel \cite{tse_book}. So the user can send a total of $B = b T_q$ feedback bits at $\epsilon(T_a,b)$ outage. Although the constellations used in practice have $2^b$ points where $b$ must be a positive integer, for the time being we relax this restriction and allow positive real values for $b$. 

We define the squared CDI error as the sine squared of the angle ($\theta$) between the true channel direction vector $\mathbf{\bar{h}}$ and the BS estimated direction vector $\mathbf{\bar{\hat{h}}}$, denoted as $\sigma^2(\mathbf{h, \hat{h}})$.
\beq
\sigma^2(\mathbf{h, \hat{h}}) \stackrel{\Delta}{=}  \sin^2(\theta) = 1 - \cos^2(\theta) = 1- |\mathbf{\bar{h}^{\dagger} \bar{\hat{h}}}|^2
\eeq
Further the MSE of CSIT is defined to be the expected value of the squared CDI error at the transmitter and denoted as $\sigma^2$. Although it's a slight abuse of notation but it has been shown that the CDI plays a vital role both for single-user and multi-user scenarios \cite{jindal_fb06}.

For the quantization of $M$-dimensional unit-norm CDI at the user, we employ random vector quantization (RVQ). For RVQ, the exact expression for the mean-square quantization error $\sigma_q^2$ has been given in \cite{love_RVQ}, \cite{jindal_fb06} as
\beq
\sigma_q^2 = 2^B \beta\left(2^B,\frac{M}{M-1}\right),
\eeq
where $B$ is the total number of feedback bits (i.e. the codebook consists of $2^B$ codes) and $\beta$ represents the beta function which is defined in terms of the Gamma function as $\beta(a,b) = \frac{\Gamma(a) \Gamma(b)} {\Gamma(a+b)}$. However it turns out that a  simple and tight upper bound given in reference \cite{jindal_fb06} suffices:
\beq
\sigma_q^2 \leq 2^{\frac{-B}{M-1}}.
\label{eq:quant_error}
\eeq
\subsection{Optimal Resource Split between Training and Quantized Feedback}
\label{sec:split_optimization}
\begin{theorem}[The minimization of the MSE of CSIT]
Under the training and feedback combining strategy, the MSE of CSIT $\sigma^2$ is minimized as a result of the following optimization governing the fixed resource ($T_{fb}$) split between the training $T_a$ and the quantized feedback interval $T_q$ and the outage rate $b$:
\beq
{\sigma^{2}}^* = \underset{T_a, b}{min} \; \left[ 2^{\frac{-b(T_{fb}-T_a)}{M-1}} + \epsilon(T_a,b) \right]
\eeq
The constraints for this minimization are:
\beq
1  \leq T_a  \leq T_{fb} \;\;\;\; \mathrm{and} \;\;\;\; 0  \leq b 
\eeq
The outage probability in the feedback interval $\epsilon(T_a,b)$ and the outage rate $b$ are linked by the relation:
\beq
b = \log \left( 1 + \frac{P^2 T_a} {2(P+PT_a+1)} F^{-1}(\epsilon(T_a,b)) \right),
\label{eq:outage_rate_and_proba}
\eeq
where $P$ is the user's power constraint and $F^{-1}(.)$ is the inverse of the standard cumulative distribution function (CDF) of $\chi^2_{2M}$ distributed variable.
\label{th:MSE_CSIT}
\end{theorem}
\begin{proof}
The proof consists of two parts. First we show the argument of minimization to be an upper bound on the MSE of CSIT and in the second part, the relation between $\epsilon(T_a,b)$ and $b$ is derived.\\
\textbf{Upper bound on the MSE of CSIT:} During the feedback phase, when the channel is not in outage and the BS is able to decode the feedback correctly, there is only quantization error in the final CSIT estimate. On the other hand, when the channel is in outage (happens with probability $\epsilon(T_a,b)$), the BS cannot decode the feedback information. Hence the MSE of CSIT $\sigma^2$ can be written as
\begin{eqnarray}
\sigma^2 & =     & \left(1-\epsilon(T_a,b)\right) \sigma_q^2 + \epsilon(T_a,b) \; \mathbb{E} \sigma^2_{\mathbf{\bar{h} \neq \bar{\hat{h}}}}(\mathbf{h, \hat{h}}) \nonumber \\
         & \leq  & \left(1-\epsilon(T_a,b)\right) \sigma_q^2 + \epsilon(T_a,b) \nonumber \\
         & \leq  & \sigma_q^2 + \epsilon(T_a,b),
\label{eq:CSIT_error}         
\end{eqnarray}
where $\sigma_q^2$ is the mean-square quantization error and $\sigma^2_{\mathbf{\bar{h} \neq \bar{\hat{h}}}}(\mathbf{h, \hat{h}})$ represents the MSE of CSIT when the channel is in outage (which means a feedback error occurs). The first inequality is obtained as $\mathbb{E} \sigma^2_{\mathbf{\bar{h} \neq \bar{\hat{h}}}}(\mathbf{h, \hat{h}})$ is upper-bounded by $1$. Putting the value of $\sigma_q^2$ from eq. (\ref{eq:quant_error}) using $B=bT_q$ and $T_{fb} = T_a + T_q$ in eq. (\ref{eq:CSIT_error}), we get the desired upper bound of the MSE of CSIT as
\beq
\sigma^2 \leq 2^{\frac{-b(T_{fb}-T_a)}{M-1}} + \epsilon(T_a,b),
\label{eq:CSIT_error2} 
\eeq
which concludes the first part of our proof.

\textbf{Significance of the MSE bound:} The MSE bound of the CSIT eq. (\ref{eq:CSIT_error2}) is the desired performance metric. Its minimization gives us the optimal values for $T_a$, $T_q$ and $b$ (the number of feedback bits per channel use - this parameter governs the constellation size and hence the quantization error) for a fixed resource $T_{fb}$. This bound shows us the basic trade-off involved. If the total number of feedback bits $B=bT_q$ is made large (either by choosing a large rate $b$ per channel use in the feedback channel or by making $T_q$ large), it will allow the user to select a larger codebook (with $2^B$ codewords) and hence the quantization error will be negligible. But this strategy will plague the final CSIT estimation error by introducing a lot of outage (due to large $b$ or poor channel estimate $\mathbf{\hat{h}_a}$ caused by small $T_a = T_{fb}-T_q$). On the other hand for a small number of total feedback bits $B$, the degradation due to outage probability will fade away, but there will be fewer codewords in the codebook and hence a large quantization error.

\textbf{The relation of $b$ and $\epsilon(T_a,b)$:} Pilot sequence transmission from the user to the BS for an interval of length $T_a$, given in eq. (\ref{eq:TDD_analog}), can be equivalently written in a simplified form as
\beq
\mathbf{y_a} = \sqrt{PT_a} \; \mathbf{h + n_a},
\eeq
where $P$ is the user's power constraint and $\mathbf{y_a, h, n_a}$ are the received signal, the channel vector and the noise respectively, all column vectors of dimension $M$. The BS can make MMSE estimate $\mathbf{\hat{h}_a}$ of the channel $\mathbf{h}$ as
\beq
\mathbf{\hat{h}_a} = \frac{\sqrt{PT_a}}{PT_a + 1} \mathbf{y_a}.
\eeq
As the i.i.d. channel entries are standard Gaussian, the MMSE estimation error $\mathbf{\tilde{h}_a = h - \hat{h}_a}$ has also Gaussian i.i.d. entries as $\mathbf{\tilde{h}_a} \sim \mathcal{CN}\left(\mathbf{0},\sigma_a^2 \mathbf{I_M}\right)$ and the MSE per channel coefficient $\sigma_a^2$ is given by
\beq
\sigma_a^2 = \frac{1}{PT_a + 1}.
\label{eq:sigma_a}
\eeq
Similarly the estimate $\mathbf{\hat{h}_a}$ has Gaussian i.i.d. entries and is distributed as $\mathbf{\hat{h}_a} \sim \mathcal{CN}\left(\mathbf{0},\frac{PT_a}{PT_a+1} \mathbf{I_M}\right)$.

Now we focus our attention on the quantized feedback interval of the CSIT acquisition, given in eq. (\ref{eq:TDD_digital}). The signal received during one symbol interval of this phase is given by
\beq
\mathbf{y_{q}} = \sqrt{P} \; \mathbf{h} x_{q} + \mathbf{n_{q}},
\eeq
where $x_{q}$ represents the scalar feedback symbol transmitted by the user and $\mathbf{y_{q}, h, n_{q}}$ are $M$-dimensional column vectors representing respectively the observed signal, the channel and the noise for this particular symbol interval. To decode this information, the BS uses the estimate $\mathbf{\hat{h}_a}$ that it developed during the training phase. So the above equation can be written as
\beq
\mathbf{y_{q}} = \sqrt{P} \; \mathbf{\hat{h}_a} x_{q} + \sqrt{P} \; \mathbf{\tilde{h}_a} x_{q} + \mathbf{n_{q}}.
\eeq
The average effective signal-to-noise-ratio (denoted as $\mathrm{SNR_{eff}}$) at the BS during the feedback interval relegating the imperfect channel estimate portion of the signal into noise and treating $\mathbf{\hat{h}_a}$ as the perfectly known channel is given by:
\beq
\mathrm{SNR_{eff}} = \frac{P ||\mathbf{\hat{h}_a}||^2}{P\sigma_a^2 + 1}.
\eeq
Plugging in the value of $\sigma_a^2$ from eq. (\ref{eq:sigma_a}), $\mathrm{SNR_{eff}}$ will become
\beq
\mathrm{SNR_{eff}} = \frac{P ||\mathbf{\hat{h}_a}||^2}{\frac{P}{PT_a+1} + 1}.
\eeq
We can do a small change of variable as $\frac{2(PT_a+1)}{PT_a} ||\mathbf{\hat{h}_a}||^2$ represents a standard chi-square random variable having $2M$ degrees of freedom (DOF), denoted as $\chi^2_{2M}$. So the $\mathrm{SNR_{eff}}$ becomes
\beq
\mathrm{SNR_{eff}} = \frac{P^2 T_a} {2(P+PT_a+1)} \chi^2_{2M}. 
\eeq

The outage probability $\epsilon(T_a,b)$ during this feedback interval corresponding to the outage rate $b$ bits per channel use can be written as
\begin{eqnarray}
\epsilon(T_a,b) & = & \mathbb{P} \left[ \log\left( 1 + \mathrm{SNR_{eff}}\right) \leq b \right] \nonumber \\
         & = & \mathbb{P} \left[ \log\left( 1 +  \frac{P^2 T_a} {2(P+PT_a+1)} \chi^2_{2M} \right) \leq b \right],
\label{eq:outage}         
\end{eqnarray}
where $\mathbb{P}$ denotes the probability of an event. This relation can be inverted to obtain the outage rate $b$ corresponding to the outage probability $\epsilon(T_a,b)$, as given below
\beq
b = \log \left( 1 + \frac{P^2 T_a} {2(P+PT_a+1)} F^{-1}(\epsilon(T_a,b)) \right),
\label{eq:outage_rate}
\eeq
where $F^{-1}(.)$ is the inverse of the CDF of $\chi^2_{2M}$ distributed variable. This concludes the proof.

The analytical solution to the minimization in Theorem \ref{th:MSE_CSIT} does not bear closed form expression but its numerical optimization is quite trivial.
\end{proof}

\section{Optimization Setup With Practical Constellations}
\label{sec:constellation_opt}
In the previous optimization procedure, we had relaxed the restriction of practical constellations and allowed any positive real values for the outage rate $b$ bits per channel use. But this is not true for the practical communication systems as the constellations used always have number of points equal to an integer power of $2$, i.e., $b$ can only take an integer value. We propose two simple strategies in the following sub-sections to handle this issue which arises due to this limitation of practical constellations.

\subsection{Resource Split Optimization for a Fixed Constellation}
\label{sec:fixed_const_opt}
We can optimize the MSE of CSIT for a fixed constellation, i.e. for a fixed outage rate $b$. In this case, the outage rate based optimization setup, built in the previous section, remains operational except that $b$ is no more an optimization variable but a fixed parameter corresponding to the chosen constellation. Thus $b$ will assume the values of $2$ and $4$ for QPSK and 16-QAM, respectively, although any other constellation can be chosen. The minimization of the MSE of CSIT will give the optimal resource split tailored for the particular constellation chosen. Hence the objective function for a fixed constellation (fixed value of $b$) becomes:
\beq
\underset{T_a}{min} \; \left[ 2^{\frac{-b(T_{fb}-T_a)}{M-1}} + \epsilon(T_a,b) \right]
\eeq
where $T_{fb} = T_a + T_q$ and $b$ are fixed, and $b$ and $\epsilon(T_a,b)$ are related as in Theorem \ref{th:MSE_CSIT}. The constraint for this minimization is:
\beq
1 \leq T_a \leq T_{fb}
\eeq
This minimization gives us the optimal value of training length $T_a$ which should be used to get the minimum MSE of CSIT for this particular constellation (fixed $b$) under fixed values of $M$, $P$ and $T_{fb}$. This restriction of fixed constellation brings in some limitations. For example, the use of smaller constellation like QSPK at very high SNR will not be beneficial as CSIT error will stay bounded due to the fixed cardinality of the codebook (hence quantization error will be non-diminishing as a function of SNR) even for asymptotically large values of SNR.
%
\subsection{Using Real Values of $b$ with Extra Parity Bits}
\label{sec:const_coding}
The other way to resolve the issue of discrete practical constellations is through the use of channel coding. This allows us to use positive real values for $b$, obtained from the original optimization setup. The only restriction, we impose, is that $B$ should take an integer value which can be obtained by using ceiling or floor operation on the product $b T_q$. Now this $B$ governs the cardinality of the codebook. The actual constellation, which is used to send feedback, is the one larger than that dictated by $b$, among the available constellations. Let the rate of that constellation be denoted by $b_c$. Hence the number of total bits, which will be sent in the feedback phase, is $B_c = b_c T_q$ where $B_c > B$ as $b_c > b$. All the extra bits $B_c - B$ in the feedback phase are used as parity bits. So one can employ either linear block codes or convolutional codes with an appropriate rate so as to convert $B$ information (true channel feedback) bits into $B_c$ coded bits. One advantage of using convolutional codes is that puncturing can give more flexibility for rate matching. Now these $B_c$ bits are sent in the digital feedback phase. As the outage rate $b$ is less than the rate $b_c$ of the constellation chosen, the use of larger constellation will give rise to increase in the number of erroneous coded bits. The number of errors will grow large in direct proportion to the difference $B_c - B$. On the other hand, all the extra feedback bits $B_c - B$ are the parity bits and when decoding will be performed at the BS, the capability of this coding/decoding operation to combat the channel errors (introduced in the quantized feedback) is also proportional to this difference, hence compensating the negative impact of using larger constellation.

\section{Simulation Results}
\label{sec:results}
Our simulation environment consists of a BS with $M=4$ antennas and a single user with a single antenna. The channel model is the same as described in Section \ref{sec:model}. The feedback interval $T_{fb}$ is fixed to $20$ channel uses for all simulations. 

\subsection{Optimization Results for Continuous Constellations}
\begin{figure}[htbp]
	\begin{center}
		\includegraphics[scale=0.9]{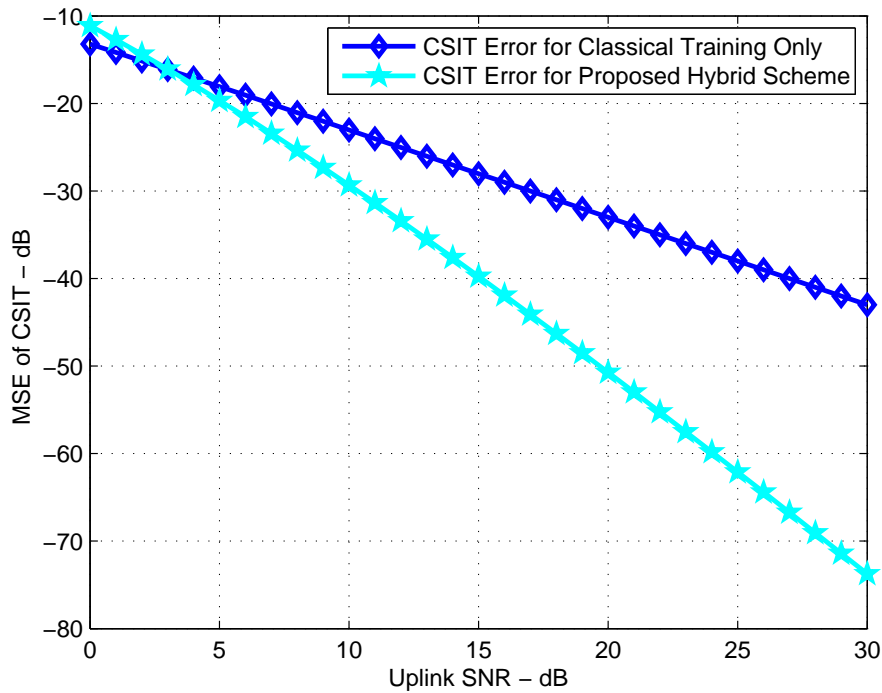}
	\end{center}
	\caption{Mean-Square CSIT Errors: $T_{fb} = 20$ and $M=4$. The novel hybrid scheme performs much better than the classical training based CSIT acquisition. Gains are significant even with naive use of practical constellations without any coding.}
	\label{fig:errors}
\end{figure}

First we present the results when the outage rate $b$ is not constrained to be an integer and can assume any positive real value. The optimization of the objective function, given in section \ref{sec:outage_based_opt}, gives us the values for the optimal training length $T_a$ and the optimal outage rate $b$ for various values of user's power constraint, which is equal to the UL SNR as the noise at every BS antenna has been normalized to have unit variance. Knowing the values of $\epsilon(T_a,b)$ and $T_q$, computed based upon the optimal values of $T_a$ and $b$, allows us to compute the upper bound of the final CSIT error eq. (\ref{eq:CSIT_error2}). These values have been plotted in dB scale in Fig. \ref{fig:errors}.
For comparison purpose, we have also plotted the MSE of CSIT with classical training based estimation. This plot clearly shows the interest for our hybrid two-staged CSIT acquisition strategy as, from medium to large SNR values, CSIT error incurred by this scheme is much less than the error obtained by training based only CSIT acquisition. Only at very low SNR values, this two stage scheme performs worse than the classical training scheme. This happens because we have restricted our final estimate to come from the digital feedback. Here the total feedback resource (SNR and $T_{fb}$) does not allow transmission of sufficient number of bits through the channel so quantization error is quite large. This gets aggravated due to the poor training based estimate based upon which these bits are decoded, further degrading the performance. This degradation can be easily avoided by selecting an SNR threshold below which traditional training based scheme should be employed.

\begin{figure}[htbp]
	\begin{center}
		\includegraphics[scale=0.9]{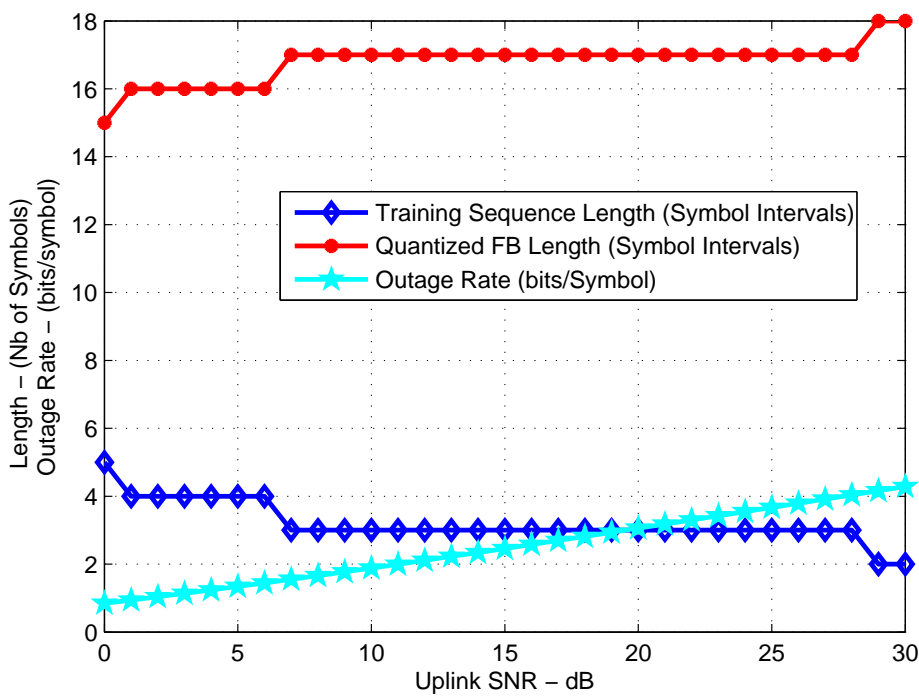}
	\end{center}
	\caption{Optimal Lengths and Outage Rate: $T_{fb} = 20$ and $M=4$. With increase in SNR, both the length of the quantized feedback interval $T_q$ and the outage rate $b$ increase gradually.}
	\label{fig:lengths}
\end{figure}

To see the optimal split between training and quantized feedback, we have plotted the optimal values of training length $T_a$, corresponding values of quantized feedback interval $T_q$ and the optimal outage rate $b$ in Fig. (\ref{fig:lengths}). 

\subsection{Optimization Results for Discrete Constellations}
\begin{figure}
   \centering
   \subfigure[ ]{
      \includegraphics[scale=0.9]{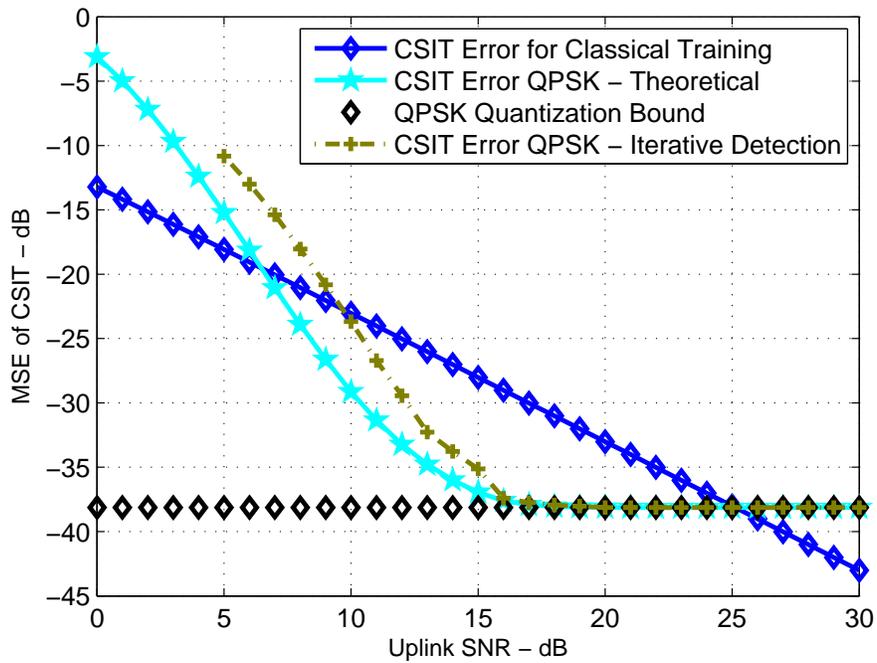}
      \label{fig:errors_QPSK}
   }
   \subfigure[ ]{
      \includegraphics[scale=0.9]{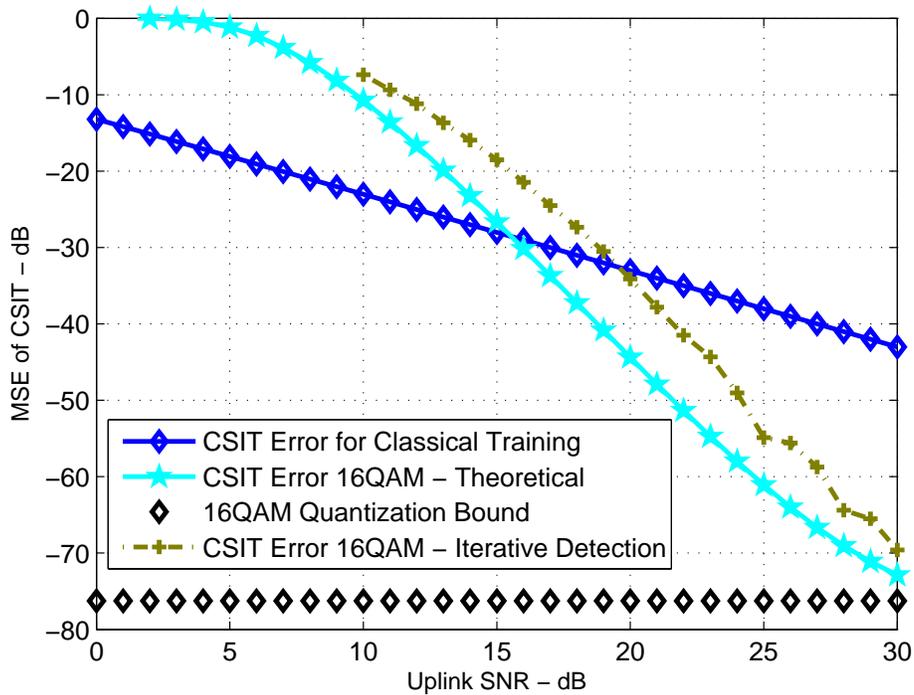}
      \label{fig:errors_16QAM}
   }
\caption{Mean-Square CSIT Errors: $T_{fb} = 20$ and $M=4$ (a) QPSK and (b) 16-QAM. The novel hybrid scheme with QPSK performs better than the classical one from $9$ to $25$ dB of SNR, but 16-QAM outperforms both after $21$ dB.}
\label{fig:errors_constell}
\end{figure}

In this section, we present simulation results when fixed constellations QPSK and 16-QAM are used for quantized feedback transmission. So the outage rate $b$ becomes fixed corresponding to the fixed constellation ($2$ for QPSK and $4$ for 16-QAM) and the optimization is carried only over the resource split between training and quantized feedback as described in \ref{sec:fixed_const_opt}. The curves for the MSE of CSIT obtained theoretically, by doing the simulations with actual constellations and the corresponding quantization bound for that constellation have been plotted in Fig. \ref{fig:errors_constell}. 
Quantization bound gives the quantization error when maximal ($T_{fb}-1$) symbols are used for quantized feedback part. Hence, it gives the lower bound on the MSE of CSIT (performance upper bound) for that particular constellation. For comparison purpose, we have also plotted the MSE of CSIT for classical training scheme. This figure shows that from low to medium SNR values, the novel scheme with QPSK gives CSIT error below that of the classical training approach but 16-QAM is not attractive in this range due to many incorrect detection events. At high SNR values, hybrid scheme with QPSK suffers from performance degradation due to its bounded quantization error but 16-QAM behaves much better than the classical scheme. At very high values of SNR, even the 16-QAM will show bounded performance for the same reason that its rate does not increase with SNR but then one needs to switch to further larger constellations.

In Fig. (\ref{fig:errors_constell}), both for QPSK and 16-QAM, we have plotted the MSE of CSIT using our proposed iterative estimation and detection algorithms from section \ref{sec:algos}. A surprising fact about the two proposed iterative algorithms is their similar performance. One would expect the iterative estimation and detection algorithm (with ML detection) to perform much better than the simplified iterative estimation and detection algorithm (which uses the simple LS detection), but extensive simulations show that the performance difference between the two algorithms is negligible. In all our simulations, both algorithms show very rapid convergence and they were always converging in second or third iteration. There were extremely rare instances (less than one in ten million) when convergence was not achieved in three iterations. 

We don't plot the optimal training and quantized feedback interval lengths out of space limitation but they show the same behavior as displayed in Fig. (\ref{fig:lengths}), i.e., the optimal quantized feedback interval gets larger with the increase in SNR for both constellations.

%

%

\subsection{Discrete Constellations and Coding}
Now we plot the results of the MSE of CSIT when quantized feedback is sent using discrete constellations and the rate matching is performed using convolutional codes as explained in section \ref{sec:const_coding}. The code rates and the puncturing patterns need to be selected carefully. First of all, convolutional codes of all desired rates are not available. Secondly, although puncturing can help a lot to reach to the desired rate still it needs to be selected carefully as random choice of puncturing pattern may destroy the code structure and hence ultimately its performance.

\begin{figure}[htbp]
	\begin{center}
		\includegraphics[scale=0.9]{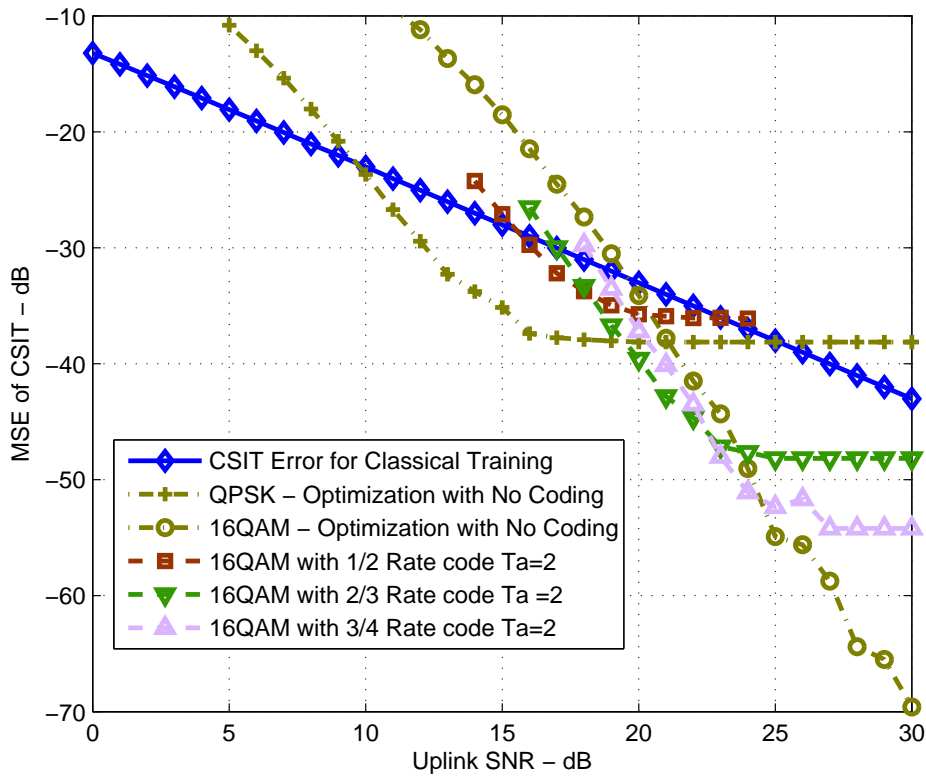}
	\end{center}
	\caption{Mean-Square CSIT Errors with Convolutional Coding: $T_{fb} = 20$ and $M=4$. At certain SNR intervals, coding strategy performs better than no coding optimal resource split outcome.}
	\label{fig:errors_code}
\end{figure}

We plot the results obtained using three different codes ($1/2$ rate code, $2/3$ rate code and $3/4$ rate code) in Fig. (\ref{fig:errors_code}). All of these codes have been used with 16-QAM ($4$ bits per channel use). Hence the number of actual information (feedback) bits are $2$, $2.67$ and $3$ per channel use for $1/2$, $2/3$ and $3/4$ rate code respectively. For comparison purpose, the plot shows the MSE of CSIT obtained by using QPSK and 16-QAM constellations without any coding and through classical training scheme. 

%
%
%
%
%
%
%
For $1/2$ rate code, the generator matrix is $[171 \; 133]_8$ and trace back length is $30$. It performs better than classical training from $16$ to $23$ dB of SNR but QPSK without any coding performs better than this curve. 
For $2/3$ rate code, the generator matrix is $[4\; 5\; 17;7\; 4\; 2]_8$ with trace back length of $20$. From $17$ dB onward, it performs better than classical training. It performs even better than 16-QAM (without coding) before $24$ dB of SNR.
For $3/4$ rate code, we use the $1/2$ rate base code (same as before) and use the puncturing pattern of $[1 1 1 0 0 1]$ to get the final rate of $3/4$.

\subsection{Imperfect CSIR Analysis}
\label{sec:results_imperfect_CSIR}
All the previous results have been obtained working under the assumption of perfect CSIR which is certainly too good to be true. Here we remove this perfect CSIR assumption and analyze how the MSE of CSIT with novel scheme behaves with imperfect CSIR.
\begin{figure}
   \centering
   \subfigure[ ]{
      \includegraphics[scale=0.9]{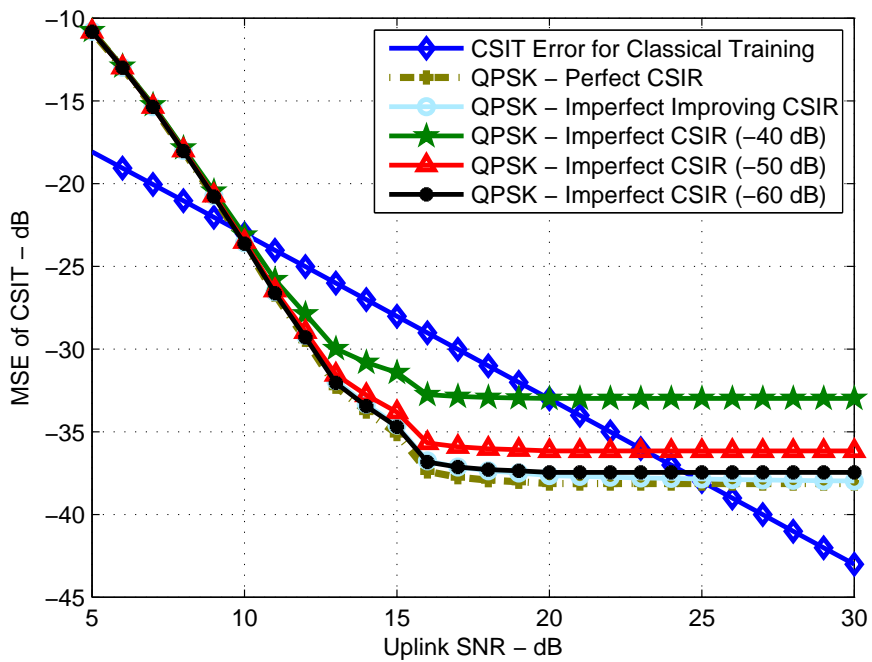}
      \label{fig:errors_QPSK_ImpCSIR}
   }
   \subfigure[ ]{
      \includegraphics[scale=0.9]{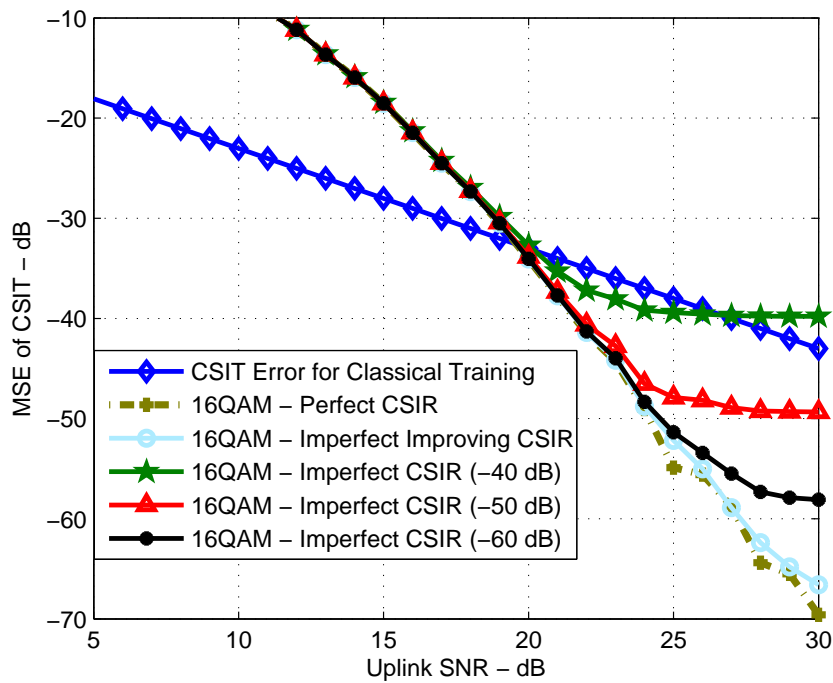}
      \label{fig:errors_16QAM_ImpCSIR}
   }
\caption{Mean-Square CSIT Errors with Imperfect CSIR: $T_{fb} = 20$ and $M=4$ (a) QPSK and (b) 16-QAM. For an imperfect CSIR of reasonable quality, the novel scheme performs much better than the classical scheme and the performance approaches to the perfect CSIR case for a good enough CSIR.}
\label{fig:errors_constell_ImpCSIR}
\end{figure}

The curves, when quantized feedback is transmitted using QPSK and 16-QAM, have been plotted in Fig. (\ref{fig:errors_constell_ImpCSIR}). We have plotted these curves under two scenarios. First, when the CSIR quality varies and improves with the increase in UL SNR which is quite logical as, due to reciprocity, the link quality improves in both directions and the BS can surely pump more power as compared to a small hand-held mobile unit. For this case, we take the MSE of CSIR $30$ dB less than the classical training only CSIT curve. The second scenario is when CSIR quality is held fixed independent of the UL SNR. For this, we plot the MSE of CSIT when the MSE of CSIR is kept fixed at $-40$, $-50$ and $-60$ dB. We believe this scenario to be of relatively less importance. We remark that when CSIR quality improves with UL SNR, hybrid approach performs very close to the perfect CSIR curve. For the other case when CSIR quality is kept fixed, it may become the performance limit of the MSE of CSIT (if not of proper quality).

\section {Concluding Remarks}
\label{sec:conc}
Traditional CSIT acquisition in reciprocal systems relying exclusively on the use of training sequences ignores the shared knowledge of an identical channel between the BS and the user. We presented a novel approach of CSIT acquisition at the BS for the DL transmission in a reciprocal MIMO communication system combining the use of a training sequence together with quantized channel feedback. We characterized the optimal CSIT acquisition setup and proposed two iterative algorithms for the resulting joint estimation and detection problem and provided a convergence proof. The novel outage-rate based approach allows the optimal resource partitioning between the training and the quantized feedback. We proposed two strategies to overcome the limitation of practical constellation availability with integer number of bits per channel use either by optimizing the resource split for a particular constellation or by the use of channel coding for rate matching. The novel combining scheme shows superior performance due to better exploitation of the reciprocity principle and the trade-off between the CSIT quality and the resource utilization improves significantly. It is further shown that with an imperfect CSIR of reasonable quality, performance gains comparable to the perfect CSIR case are achievable.\\
\textbf{Multi-User Extension:} The proposed novel scheme holds verbatim in the case of multiple users. In the first phase of ``pure training", the users should use orthogonal training signals so that the BS gets an initial estimate of the channel. Then during the second ``quantized feedback" phase, the UL channel should be used as MIMO-MAC. The optimization of resources remains however an open problem in this setting. In this scenario, the resource optimization will depend heavily upon the BS transmission strategy, e.g., the optimal resource split could be extremely different for TDMA or SDMA. The presence of more users in the system, larger than the BS transmit antennas, and subsequently required user scheduling would add an extra twist to this problem.\\
\textbf{Users with Multiple Antennas:} There are different ways to treat the fully general case of multiple users with multiple antennas where even a single user can be transmitted multiple streams. It adds an extra level of complexity to the open problem of multiple single-antenna users. For the users with multiple antennas, a simplifying strategy could be to do antenna combining as in \cite{jindal_AntComb} to minimize the quantization error. This scheme is promising as it reduces the feedback requirement by converting the MIMO channel into a vector channel and in a direction of minimal quantization error. Hence effectively it will become the multiple single-antenna user extension of our work.

\bibliographystyle{IEEEbib}
\bibliography{strings,refs}

\end{document}